\documentstyle[prc,aps,epsf]{revtex}

\begin{document}

\input epsf

\newcommand{\be}{\begin{equation}}
\newcommand{\ee}{\end{equation}}
\newcommand{\beqn}{\begin{eqnarray}}
\newcommand{\eeqn}{\end{eqnarray}}

\title{KEWPIE: a dynamical cascade code for 
decaying exited compound nuclei}
\author{Bertrand Bouriquet$^a$, Yasuhisa Abe$^a$ and David 
Boilley$^{a,b}$}
\address{$^a$ Yukawa Institute for Theoretical Physics, Kyoto 
University,
Kyoto 606-8502, Japan\\
$^b$ GANIL, BP 55027, 14076 Caen cedex 05, France}

\date{Submitted to Comp. Phys. Com. on 22nd  July 2003}
\maketitle

\begin{abstract}
A new dynamical cascade code for decaying hot nuclei is proposed and 
specially adapted to the synthesis of super-heavy nuclei. For such a 
case, the interesting channel is the tiny fraction that will decay 
through particles emission, thus the code avoids classical Monte-Carlo 
methods and proposes a new numerical scheme.
The time dependence is explicitely taken into account in order to 
cope with the fact that fission decay rate might not be constant. The 
code allows to evaluate both statistical and dynamical observables.
 Results are successfully compared to experimental data.
\end{abstract}
\vspace{0.5cm}


\section{Introduction}
Since the pioneering work of Bateman \cite{Bat} who developed a 
general
equation for radioactive decay chains, the model has found many
 applications in nuclear physics and in other fields. Purely
statistical version of the problem, when the time is integrated out,
has been extended to multi-channel cascade process of decay of  
exited compound 
nuclei that can evaporate neutrons, protons, alphas\ldots\ and
successfully compared to experimental data. The original time dependent 
Bateman equations can also
easily be extended to the multi-channel process, provided that the
decay widths are time-independent.  However, 
the decay width of the fission channel is now known not to be 
time-independent,
so  Bateman problem should be solved numerically. 
Actually in fission of excited compound nuclei, 
it is known that it takes a finite time to build up the
quasi-stationary probability flow over the fission barrier. 
Therefore, the time evolution should be taken into account. For recent
reviews on the dynamics of fission, see Refs. \cite{Hil,Abe}.

One of our purposes is to apply the method to evaluate the 
 production cross section of the superheavy elements. In such 
reactions, most 
of the compound nuclei will undergo fission, and thus we need to 
evaluate 
the very tiny fraction that will decay only through particles 
emission. 
Therefore, classical Monte-Carlo methods \cite{Gav,Cole} are not 
suitable. We propose here another numerical scheme. Furthermore  
we construct a new statistical  code named 
KEWPIE (Kyoto Evaporation Width calculation Program with tIme 
Evolution).

Fusion reactions that lead to the production of super-heavy elements 
involve often a combination of rather heavy ions for the entrance 
channel where it is well known that fusion is strongly hindered. The 
extra push parametrisation gives only a qualitative behaviour of this 
fusion hindrance, so more precise calculation should be done. There 
are several way of taking  into account the hindrance,  one of which 
is to use the two step model \cite{Abe2,Shen,Kos}. The   code is 
designed to accommodate fusion  probabilities calculated by such a 
model, in addition to the probabilities given as usual by 
transmission coefficients whose calculation with the full proximity 
potential is included in the code as standard.

\section{Dynamical framework}

\subsection{Single chain}
\subsubsection{Populations as a function of time}
Before generalising the method to the multi-channel decay problem, we
first recall Bateman's results considering a single chain starting 
from a single nucleus. The equations then
read,
\beqn
\frac{dP_1}{dt}&=&-\Gamma^t_1 P_1 \quad {\rm with} \quad P_1(0)=1,\\
\frac{dP_2}{dt}&=&\Gamma_1P_1-\Gamma^t_2 P_2 \quad {\rm with} \quad 
P_2(0)=0\\
\frac{dP_3}{dt}&=&\Gamma_2P_2-\Gamma^t_3 P_3 \quad {\rm with} \quad
P_3(0)=0\\
&\vdots&\nonumber\\
\frac{dP_n}{dt}&=&\Gamma_{n-1}P_{n-1}-\Gamma^t_n P_n \quad {\rm with} 
\quad
P_n(0)=0,
\eeqn
where $\Gamma^t_{n}$ is the total decay width for the nucleus $n$ and 
$\Gamma_{n}$ is the evaporation width, whatever the particle is. 
$P_{n}(t)$ denotes the population of the nucleus $n$ at time $t$. 
If the decaying widths are time independent, it is very easy to solve 
analytically these coupled differential equations,  by
taking the Laplace transform as Bateman did. This leads to
\beqn
\tilde{P}_1(s) &=& \frac1{\Gamma^t_1+s}\\
\tilde{P}_2(s) &=& \frac1{\Gamma^t_2+s}\frac{\Gamma_1}{\Gamma^t_1+s}\\
&\vdots&\nonumber\\
\tilde{P}_n(s) &=& \frac1{\Gamma^t_{n}+s} \prod_{i=1}^{n-1}
\frac{\Gamma_i}{\Gamma^t_i+s},
\eeqn
where $\tilde{P}(s)$ is the Laplace transform of $P(t)$.
The
solution is then obtained by taking the inverse Laplace transform,
\beqn
P_1(t)&=&e^{-\Gamma^t_1t}\\
P_2(t)&=&\frac{\Gamma_1}{\Gamma^t_2-\Gamma^t_1}
(e^{-\Gamma^t_1t}-e^{-\Gamma^t_2t})\\
&\vdots&\nonumber\\
P_n(t)&=&\prod_{k=1}^{n-1} \Gamma_k \sum_{i=1}^n 
\frac{e^{-\Gamma^t_it}}{\prod_{j\neq 
i}(\Gamma^t_j-\Gamma^t_i)},\label{Pn}
\eeqn
where we have supposed that the $\Gamma^t_n$'s are all different from
each other. A special attention should be drawn on the last nucleus of
the chain, for which the solution might differ from the above when
$\Gamma^t_{n_{max}}=0$. For such a case,
\beqn
P_{n_{max}}&=& \Gamma_{n_{max}-1}\int_0^{+\infty}dt\,
P_{n_{max}-1}(t)\\
&=& \prod_{k=1}^{n_{max}-1} \Gamma_k \sum_{i=1}^{n_{max}} 
\frac{1-e^{-\Gamma^t_it}}{\Gamma^t_i\prod_{j\neq 
i}(\Gamma^t_j-\Gamma^t_i)}.
\eeqn

From
the populations, we can then calculate any observable. This solution 
is only valid for time independent $\Gamma^t$'s and then for a given 
excitation energy for each nucleus. It is then not always 
applicable, but it will be very useful to test our numerical scheme.

\subsubsection{Statistical observables}
Assuming that only neutrons are evaporated along the chain, 
the total width for the isotope $n$ that can also undergo fission, is then 
$\Gamma^t_{n}=\Gamma^f_{n}+\Gamma_{n}$, where $\Gamma^f_{n}$ 
corresponds to the 
fission width. The
probability to emit exactly $n$ neutrons prior fission is 
\beqn
p_{n-1}&=&\int_0^{+\infty}dt\, \Gamma^f_n P_{n-1}(t)\\
&=&\frac{\Gamma^f_n}{\Gamma^t_n}\prod_{i=1}^{n-1} 
\frac{\Gamma_i}{\Gamma^t_i}
\eeqn
which is in agreement with the reults of Ref. \cite{Has}. For the end of the chain, 
\be
p_{n_{max}}=\prod_{i=1}^{n_{max}-1} \frac{\Gamma_i}{\Gamma^t_i}.
\ee
These results are commonly used in statistical decay codes. The
average pre-scission neutron multiplicity is simply,
\be
\langle n\rangle= \sum_{n=1}^{n_{max}} (n-1)\,p_n.
\ee

The total number of
nuclei that fission is that $N_f=1$ if $\Gamma_{n_{max}}\neq0$ and
$N_f=1-p_{n_{max}}$ else.

\subsubsection{Fission time}\label{sft}
The average fission time is an interesting observable to test the dynamics of
the cascade because it can be measured directly in some cases. We 
define it as 
\beqn
\tau & = & - \int_0^{+\infty} t \frac{dP_{tot}}{dt} dt \\
& = &  \sum_{n=1}^{n_{max}} \int_0^{+\infty} t \Gamma_n^f P_n(t) dt
\label{fisti}
\eeqn

\subsection{Multichannel scheme}

Excited compound nuclei can evaporate other particles than neutrons. 
Protons,
alphas should also be taken into consideration. The populations will
be labelled as $P_{i,j}$ with $i$ being the number of evaporated
neutrons and $j$ the number of evaporated protons. In order to keep a
triangular shape of the matrix, the populations will be ordered
following the number of evaporated nucleons, starting with the
neutrons. Or, in a more formal way, $n=\frac{(i+j)(i+j+1)}2+j$ where
$n$ is a ranking number for each nucleus produced in the decay 
process.

Then, similar differential equations can be written, taking into
account, neutrons, protons, alpha evaporation and fission. The first
ones read
\beqn
\frac{dP_{0,0}}{dt}&=&-\Gamma^t_1 P_{0,0} \quad {\rm with} \quad 
P_{0,0}(0)=1,\\
\frac{dP_{1,0}}{dt}&=&\Gamma^n_{0,0}P_{0,0}-\Gamma^t_{1,0} P_{1,0} 
\quad {\rm with} \quad P_{1,0}(0)=0\\
\frac{dP_{0,1}}{dt}&=&\Gamma^p_{0,0}P_{0,0}-\Gamma^t_{0,1} P_{0,1} 
\quad {\rm with} \quad P_{0,1}(0)=0\\
\frac{dP_{2,0}}{dt}&=&\Gamma^n_{1,0}P_{1,0}-\Gamma^t_{2,0} P_{2,0} 
\quad {\rm with} \quad P_{2,0}(0)=0\\
\frac{dP_{1,1}}{dt}&=&\Gamma^n_{0,1}P_{0,1}+\Gamma^p_{1,0}P_{1,0}
-\Gamma^t_{1,1} P_{1,1} \quad {\rm with} \quad 
P_{0,1}(0)=0\label{mix}\\
\frac{dP_{0,2}}{dt}&=&\Gamma^p_{0,1}P_{0,1}-\Gamma^t_{0,2} P_{0,2} 
\quad {\rm with} \quad P_{0,2}(0)=0\\
&\vdots&\nonumber\\
\frac{dP_{2,2}}{dt}&=&\Gamma^p_{2,1}P_{2,1}+\Gamma^n_{1,2}P_{1,2}
+\Gamma^{\alpha}_{0,0}P_{0,0}-\Gamma^t_{2,2} P_{2,2} 
\quad {\rm with} \quad P_{2,2}(0)=0\\
&\vdots&\nonumber
\eeqn
Here $\Gamma^t_{ij}$ corresponds to the total decay width of the 
nucleus $ij$, including fission. The other $\Gamma$'s correspond to 
evaporation widths.
The Laplace transform of the first equations is very similar to what
was done with a single chain, up to Eq. (\ref{mix}) for which some
differences appear:
\be
\tilde{P}_{1,1}(s)=\frac{\Gamma^n_{0,1}\Gamma^p_{0,0}}{(\Gamma^t_{1,1}+s)
(\Gamma^t_{0,1}+s)(\Gamma^t_{0,0}+s)} + 
\frac{\Gamma^p_{1,0}\Gamma^n_{0,0}}{(\Gamma^t_{1,1}+s)
(\Gamma^t_{1,0}+s)(\Gamma^t_{0,0}+s)}.
\ee
The result corresponds to the sum of single chains over all possible
paths. Just considering neutrons and protons evaporation, there are
$\frac{i!j!}{(i+j)!}$ possible paths from the nucleus $(0,0)$ to the
nucleus $(i,j)$. Then, taking the inverse Laplace transform, one ends
up with a general formula that is of course the sum of single chain
terms over all possible paths. This general property due to the
linearity of the equations is also true for the observables, for which
time integration should be carried out. It is possible to recollect 
similar
terms together, but it will make formulae more complicated.

The complexity of the problem is eventually due to the multi-channel
scheme as in any cascade code, not to the dynamics that could be 
easily
added exactly to a statistical code, assuming that the $\Gamma$'s are 
time independent.  Nevertheless, 
these equations have the merit to be exact, but based on an 
approximate average model. Since we are not only interested in the 
main stream decay channel but also small decay paths, a numerical 
solution is necessary. In addition, this model does take into 
consideration $\gamma$ emissions that cool down nuclei. 
These results will then be useful for testing numerical scheme and 
understand physical results.

\subsection{Numerical scheme} \label{nusc}

For the sake of simplicity, we will only describe the scheme for a 
single 
chain. It could be easily extended to a multi-channel cascade.  
In the code each time step is larger than the previous one: 
$\delta t_{i+1}= \alpha \delta t_i$.  The alpha parameter permits 
to have an increasing time step along the cascade, and thus permits 
to take into account processes that can have very different time 
scales. The parameter $\alpha$ is usually set to $1.2$ which permits 
to cover a wide range of excitation energy. Larger values of the 
parameter $\alpha$ may induce a divergence in the fission time 
calculation (equation  \ref{fisti}) due to the multiplication of a 
small probability by a long time. The initial value $\delta t_0$ is 
set to $1 \hbar / MeV$. This choice is made so that  initial time 
scale is longer than the mean collision time (around $0.1 \hbar 
/MeV$) and shorter than the mean evaporation time (few tens of $\hbar 
/MeV$). This value is physically reasonable for a wide range of 
application. 

During a time $\delta t_i$, the population of a given nucleus $P_{n+1}$ 
increases by the decay of a mother nucleus (if exits).  At a given 
excitation energy $E^*_{n}$, this feeding 
term is $ P_{n}(t_i) N_{n}^i $ with
\be
N_{n}^i = 
\frac{\Gamma_{n}(E^*_{n})}{\Gamma_{n}^t} (1-e^{-\Gamma_{n}^t 
\delta t_i}).
\ee
During the same time this population is 
decreasing  
according to an exponential law : $P_{n+1}(t_i+\delta t_i) = P_{n+1}(t_i) 
D_{n+1}^i$ with  $D_{n+1}^i =  e^{-\Gamma_{n+1}^t \delta t_i}$. Finally the 
iterative equation for the population $P_n$ reads, 
\be \label{dec}
P_{n+1}(t_i + \delta t_i) =  P_{n+1}(t_i) D_{n+1}^i +  P_{n}(t_i) N_{n}^i.
\ee
This is for an averaged excitation energy.

In the multichannel scheme the representation is more complicated due 
to the interconnection between the  different ways of 
disintegrations. However, the problem can be treated in the same 
framework. We can calculate the number of nuclei going to fission at 
each step by looking at the quantity  $ P_{n-1}(t_i) F_n^i$  with  
$F_{n-1}^i = \frac{\Gamma^f_n}{\Gamma_n^t} (1-e^{-\Gamma_n^t \delta 
t_i})$ for each nucleus $n$ at the time $t_i$. The fission time can 
easily be deduced with a discretised version of  equation 
(\ref{fisti}). With this framework we can solve more sophisticated 
dynamics including cooling by  gamma emission or a transient time during 
which fission probability is reduced. 

\subsection{Energy spectra}

One of the particularity of the KEWPIE code is that the energy 
spectra of the produced nuclei are completely calculated and 
processed. Monte-Carlo methods were not considered because we are 
interested in the tiny fraction of events leading to super-heavy 
elements. For a given mother nucleus $n$ with a given excitation 
energy $E^*_n$, the energy spectrum of the daughter nucleus $n+1$ is 
proportionnal to $\gamma(E^*_n\rightarrow E^*_{n+1})$ entering the 
evaporation width, 
\be \Gamma_n(E^*_n)=  \int_{V_{coul}}^{E^*_n} d\varepsilon\, 
\gamma(E^*_n\rightarrow E^*_{n+1}),
\ee
where $\varepsilon=E^*_n-E^*_{n+1}$ is the energy taken by the 
evaporated particle and $V_{coul}$ is the Coulomb barrier between the daughter nucleus 
4and the evaporated particle. The former is naturally zero for neutrons. 
$\gamma(E^*_n\rightarrow E^*_{n+1})$ is the partial width for a given value of $\varepsilon$,
\be
 \gamma(E^*_n\rightarrow E^*_{n+1}) \propto 
 \frac{\rho(E^*_{n+1})}{\rho(E^*_n)}
\ee
where $\rho$ represent the level of density. The detailed formula is given in section \ref{evap}.
When the mother nucleus has itself an energy spectrum $S_n(E^*_n,t)$ 
normalized to the population, $\int_0^{+\infty}S_n(E^*_n,t)\,dE^*_n = P_n(t)$, 
the energy spectrum of the daughter is feeded by
\be
\delta S_{n+1}(E^*_{n+1},t_i+\delta t_i) = \int_0^{+\infty} dE^*_n\, S_n(E^*_n,t_i) 
\frac{\gamma(E^*_n\rightarrow E^*_{n+1})}{\Gamma_n^t}(1-e^{-\Gamma_n^t 
\delta t_i}).
\ee
In this equation, the total width $\Gamma^t_n$ is already integrated 
over the excitation energy $E^*_n$. 
The spectrum of the daughter nucleus will be used as an input for the next step 
of the cascade. The spectrum of the mother nucleus is modified 
with respect to the decay toward the daugther.
Practically, in the code, the energy spectrum is evaluated by discretizing the 
previous integral in $N$ excitation energy bins.

\section{Physical ingredients}

The physics of the problem is then in the various widths entering in 
the previous equations. In this part we will specify our choice for 
super-heavy elements. In the code the integrals are evaluated by 
discretizing the sum.

\subsection{Particle evaporation width}{\label{evap}}
Evaporation process in excited compound nuclei was modelised a 
very long time ago by Weisskopft \cite{Wei} with the principle of 
the detailed balance.
 But this modelisation does not 
take into account the angular momentum explicitly 
though it is a very sensitive 
parameter to modelise the competition between fusion and evaporation.  
We prefer the Hauser-Feschbach 
approach \cite{Haufes} which gives the width of the disintegration of 
a compound nucleus $C$ at
excitation energy $E^*_C$ and spin $J$ toward a nucleus $B$ by the 
emission of a particle $b$ of spin $I_b$,
\be
\Gamma^J_{ \{C;J;E^*_C\} \rightarrow \{ b;I_b\} }  = \frac{1}{2 \pi 
\rho^{g}_C(E^*_C ; J)}  \sum_{I_B} (2I_b+1)(2I_B+1) 
\int_{V_{coul}}^{E_C^* 
- B_b} {d \varepsilon_b} \rho^{g}_B(E^*_C - B_b - \varepsilon_b ; 
I_B)   
\sum_{S_b = |I_b - I_B |}^{I_b + I_B} \sum_{l_B = |J-S_b|}^{J+S_b}  
{\cal T}^J_{l_b S_b}(b,\epsilon_b).
\ee
In this equation $\rho^g$ 
represents the level density at the ground state and  ${\cal 
T}^J_{l_b S_b}$ the 
transmission coefficient. Detailed calculations of those coefficients 
are 
described in section \ref{ssld} and \ref{sstc}, respectively. The 
binding energy of the particle $b$ in the nucleus $B$ is represented 
by $B_b$. The variable $\varepsilon_b $ is the kinetic energy of the 
evaporated particle $b$. To obtain the total width, a 
summation over all the kinetic energies of the emitted particles is 
done from the Coulomb barrier between the 
nucleus $B$ and $b$,  named  $V_{coul}$, to the maximal 
value reachable $E^* - B_b$.

We will use in the remaining part of this publication a simpler 
notation $\Gamma^J_{\xi}$  in place of  $\Gamma^J_{ \{C;J;E^*_C\} 
\rightarrow \{ b;I_b\} 
}$ where $\xi $ is the evaporated particle.

\subsection{Fission width}{\label{sscfw}}
In order to calculate the fission width, we need to know the height 
of the fission barrier $B_{f}$ which is calculated in  detailed in 
section \ref{ssfisbar}. Then the width is given by the Bohr formula 
\cite{Bowe}, in which  a transmission coefficient factor ${\cal 
T}_f(\varepsilon_f)$ is included, 
\be
\Gamma^{BW}_{ \{C;J;E^*_C \} \rightarrow \{ f;J;E_f \} } = \frac{1}{2 
\pi 
\rho^{g}(E^*_C;J)} \int_{-B_f}^{E^*_C - B_f} d \varepsilon_f 
\rho^{s}(E^*_C - B_f-\varepsilon_f;J) {\cal T}_f(\varepsilon_f).
\ee
In this equation $\rho^g$ and $\rho^s$ represent the level densities 
of the nucleus at the ground  and saddle points, respectively, see 
section \ref{ssld}. 
The transmission coefficient is given by the Hill and Wheeler formula 
\cite{Hiwe}. This factor takes into account the effect of sub-barrier 
fission,
 
\be
{\cal T}_f(\varepsilon_f) = \frac{1}{1+e^{ \left (- \frac{2 \pi 
\varepsilon_f}{\hbar \omega_{sd}} \right ) }},
\ee
where the parameter $\omega_{sd}$ is related to the curvature of the 
fission barrier at the saddle point. 
We can define in the same way  $\omega_{gs}$ as the curvature of this 
potential for the deformation around the ground state. In this code   
$\hbar \omega_{gs}$ and   $\hbar \omega_{sd}$ are usually set to $1 
MeV$.  The variables $\omega_{sd}$ and $\omega_{gs}$ are also useful 
in order to add the Kramers \cite{Kram} and the Strutinsky 
\cite{Stru} correction factors to the fission width :

\be \label{fw}
\Gamma_{ \{C;J;E^*_C \} \rightarrow \{ f;J;E_f \} }^{K} = {\cal K}   
{\cal S}  \Gamma^{BW}_{ \{C;J;E^*_C \} \rightarrow \{ f;J;E_f \} }.
\ee

The Strutinsky factor $\cal S$ is given by

\be
{\cal S} =  \frac{\hbar \omega_{gs}}{T}.
\ee
This pre-exponential factor includes temperature $T$ ($E^* = a T^2$) 
which cancels out the temperature dependence implicitly included in 
Bohr-Wheeler formula.
 
The Kramers factor $\cal K$ reads, 

\be \label{krame}
{\cal K}  = \sqrt{1 + x_k^2} - x_k ,
\ee

with 

\be
x_k =  \frac{\beta_f}{2 \omega_{sd}}.
\ee

Those formulae take into account the effect of the viscosity of the 
nuclear matter against the fission process, where $\beta_f$ is the 
reduced friction coefficient {\it i.e.} the friction coefficient 
divided by the inertia mass. Its value is not well known yet and  set 
around $10^{19} s^{-1}  $ to  $10^{21} s^{-1}$.   

The notation for the width given in equation (\ref{fw}) 
will be simplified as $\Gamma^J_{f}$.

\subsection{Gamma emission width}

Another mode of de-excitation of the compound nucleus is the 
gamma-ray 
emission. This process reduces the excitation energy of the nucleus 
without changing the nucleus itself. That de-excitation can lead the 
nucleus energetically under the other thresholds of disintegrations 
(fission and evaporation) and so contributes to the stabilisation of 
the excited compound nucleus. The width of the gamma emission is 
governed
by the following formula \cite{Grgi,Rud}
\be
\Gamma_{ \{C;J;E^*_C \} \rightarrow \gamma} =  \frac{1}{2 \pi 
\rho(E^*_C;J)}
\sum_L \sum_{I=|J-L|}^{J+L} \int_{E^L_{\gamma m}}^{ E^L_{\gamma M}}  
d \varepsilon_{\gamma} \rho_{C}(E^*_C - \varepsilon_{\gamma};I) \xi_L 
\varepsilon_{\gamma}^{2L+1}.
\ee

In the calculation code KEWPIE  we only consider the  transitions 
with L=1 and L=2.  Integration is done between a minimal 
$E^L_{\gamma m}$  and maximal $E^L_{\gamma M}$  allowed for the 
energy of gammas : 
$E^1_{\gamma m} = 1$ MeV and  $E^1_{\gamma M } = 7$ MeV for $L=1$ 
and  $E^2_{\gamma m} = 1$ MeV and  $E^2_{\gamma M} = 4$ MeV for $L=2$.
The factors  $\xi_1$ and $\xi_2$ are parametrised as follow
\beqn
\xi_1 &=& 4.24 ~ 10^{-9} A^{\frac{2}{3}} \\
\xi_2 &=& 3.00 ~ 10^{-12} A^{\frac{5}{3}} .
\eeqn
Theses values only depend on the mass of the nucleus $A$.

\subsection{Mass and Shell correction}{\label{msc}}

Values of nuclear mass $M_A$ and  shell corrections $\Delta 
E_{Shell}$ of 
nuclei are taken from the M\o ller {\it et al} mass table 
\cite{Moll}. 
In this table are 
referenced both experimental and theoretical values for mass and 
shell correction. If the experimental value is known, this value is 
taken. In the other cases the masses calculated by the M\o ller
are used, which include macroscopic liquid drop energies and  
 the shell correction energies obtained by
Strutinsky method.  
Another alternative is to use the mass table proposed by Koura 
\cite{Kou}.

\subsection{Fissility}

The fissility of nucleus is  used as an index to evaluate many 
quantities. 
It is defined by the Coulomb energy divided by twice of the surface 
energy of the nucleus and can be given  practically in 
several ways. The simplest standard definition is $x =Z^2/50A$, with 
$A$ and $Z$ representing the mass and charge 
numbers of the nucleus, respectively.
 For heavy nuclei, we will prefer another  
parametrisation taking into account the effect of isospin. The ratio 
between neutrons and protons changes the sensitivity of the nuclei to 
the fission process. More over this formula of the fissility is  
deeply 
related to an evaluation of the fission barrier that  is described in 
the next section. A formula well  
adapted to massive systems is  given by  K. H. Schmidt {\it et al.} 
\cite{Fiss} which is used in the present code,
\be
x = \frac{Z^2}{49.22A(1-0.3803I^2)-20.489I^4)},
\ee
where $I = N-Z = A-2Z$ is the isospin.  We principally use this 
formula that, as mentioned before, is used for the parametrisation  
of the liquide drop part of the fission barrier \ref{ssfisbar} in order  to keep a 
coherence between the various parts of the code.

\subsection{Fission barrier}{\label{ssfisbar}}

There are many ways to parametrise the fission barrier $B_{f0}$, the 
Liquid Drop Model (LDM) fission barrier for $J = 0 $. 
The code is basically  designed to use two parametrisations.
The first one is the classical  Cohen-Plassil-Swiatecki (CPS) 
\cite{Cps} parametrisation that is multiparametric fit of the height 
of the fission barrier as a function of deformation of the nucleus 
based on a liquid drop model. 

As the purpose of the KEWPIE code is mainly to deal with heavy and 
super-heavy elements, another parametrisation is also implemented in 
the code. 
In the paper related to heavy elements, K.H. Schmidt{\it et al} 
\cite{Fiss} proposed a  phenomenological prametrisation of the liquid 
drop model fission 
barrier as a function of the fissility. This parametrisation is 
obtained 
 by adjusting parameters with data on 
a wide range of heavy nuclei compiled by Dahlinger {\it et al.} 
\cite{Dahl}. The results of the fitting permit to reproduce the data 
related to the disintegration of the 
 heavy elements  relatively well over a wide range of nuclei. 
Then with this parametrisation, in the 
code, the fission barrier is taken as,
\be 
B_{f0} =  \frac{0.7332 \, 10^{u} Z^2}{A^{\frac{1}{3}}},
\ee
with $u =  0.368-5.057x+8.93x^2 -8.71x^3$ where $x$ is the fissility 
of the  nucleus under consideration.
The formula represents the height of the barrier for a null angular 
momentum and without taking into account the shell correction energy 
$\Delta E_{Shell}$. Thus it has to be modified  to take in account 
those effects. 
First, the evolution from the spherical shape at the 
ground state to the deformed state at the saddle point leads to a 
modification of the barrier due to the change of rotational energy:
\be
B_{f1} = B_{f0} + (E^{saddle}_{rot} - E_{rot}^{ground}),
\ee 
where $E^{saddle}_{rot}$  and $E_{rot}^{ground}$  represent the 
rotational energies at the saddle point and the ground state, 
respectively.  

Subsequently the shell 
energy correction $\Delta E_{shell}$ mentioned in section  
\ref{ssfisbar} is substacted to obtain the true 
value of the height of the fission barrier, as
\be
B_f = B_{f1} - \Delta E_{shell}.
\ee

We are assuming that the shell correction at the saddle point can be 
neglected.
This is the final value of the fission barrier $B_f$ that is used 
in the calculation of the fission width \ref{msc}.

\subsection{Level density}{\label{ssld}}
In the program KEWPIE the evaporation of particles is calculated for 
each partial cross section as a function of the angular momentum $J$. 
So we need a level density function with angular momentum specified 
and 
we take one given in the textbook by  Bohr 
and Mottelson \cite{Bomo1},
\be 
\rho(E^*,J) = \frac{\pi^{\frac{1}{6}}}{48}6^{-\frac{1}{3}}(2J+1) a 
{\left( 
\frac{a}{2 \Im} \right) }^{\frac{2}{3}} \frac{ e^{2 \sqrt{ 
{a (E^* - E_{rot})}  }} }{(a(E^* - 
E_{rot}))^{\frac{7}{4} }}.
\ee
The rotational energy of the nucleus is given by
\be
E_{rot} = \frac{J(J+1) \hbar^2}{2 \Im }.
\ee
In this formula $J$ represents the angular momentum quantum number 
of the nucleus. To 
obtain the rotational energy we need to calculate the 
moment of inertia $\Im$ which is detailed in the next
section. 

The level density is the very key physical quantity in the 
calculations because 
it comes in all the disintegration widths, whatever 
the processes are, {\it i.e.} , evaporation, fission and gamma 
emission. 

\subsection{Moment of inertia}{\label{ssinmo}}

In the program KEWPIE we will assume that the nucleus at the ground 
state is spherical. It is reasonable for superheavy 
elements in the sense that the most stable nuclei that we hope to 
form are magical nuclei and neighbours.
 Since their fissility is high, the nuclei 
are not very deformed also at the saddle, 
so the evolution from the ground state to the saddle point is rather 
small. 
The moment  of inertia 
for the spherical nucleus is evaluated as a rigid body  
\cite{Bomo2} as follow,
\be 
\Im_{ground} = \frac{2 A r_A^2}{5 \hbar^2 }.
\ee 
The moment of inertia  at the saddle point takes into account the 
deformation of the nucleus by multiplying the momentum at  the  
ground state by a modification factor given by Hasse and Myers 
\cite{Hasmy},
\be \label{isa}
\Im_{saddle} = \Im_{ground} \left( 1 + \frac{7}{6}y \left( 1 + 
\frac{1396}{255}y    \right)    \right).
\ee
In this equation $y = 1 - x$ with $x$ the fissility of the nucleus. 
We have limited this expansion to the second order in the code, as it 
is suggested in reference \cite{Hasmy}. Keeping this limitation in 
equation (\ref{isa}) at the second order in $y$ we can express 
$\Im_{saddle}$  as a second order function of $\alpha_2$ 
(factor of the second order Legendre polynomial development of the nucleus shape) 
by the following formula,

\be 
\Im_{saddle} = \Im_{ground} \left( 1 + \frac{1}{2} \alpha_2   + 
\frac{424}{85}  \alpha_2^2    \right).
\ee

For the level of density at the ground state or the saddle point  we 
will use the corresponding  parametrisation for moment of inertia.  
The saddle point deformation can be determined by shell correction 
energy. In that case, it  can be read from a file  in order to 
accommodate results  of other microscopic calculations.  

\subsection{Level density parameter}{\label{ssldp}}

In this program the level density parameter $a$ is given by the 
formula of T\"oke and Swiatecki \cite{Tosw},
\be
a = \frac{A}{14.61}(1+3.114{\frac{F_2}{A^{\frac{1}{3}}}} + 
5.626{\frac{F_3}{A^{\frac{2}{3}}}}  ),
\ee
where $F_1$ and $F_2$ 
are related to the deformation of the nucleus as the 
variation of the surface energy and curvature energy, respectively. 
At the ground state, $F_2$ and $F_3$ are
equal to $1$, as we assume no deformation of the 
nucleus. To take into account the effects of deformation at the 
saddle 
point, we use again the 
parametrisation done by Hasse and Meyer \cite{Hasmy},
\beqn
F_2 &=&  1 + \frac{98}{45}y^2 \left( 1 - \frac{974}{765}y    
\right),  \\
F_3 &=&  1 + \frac{98}{45}y^2 \left( 1 - \frac{124}{765}y    \right) 
. 
\eeqn
where we have again limited the expansion to the second order. 
In those relations  $y = 1- x$ which is expected to be very small 
for the heavy 
or super heavy nucleus. 

At the ground state, an effect of the energy-dependence of the shell 
correction is taken into account, according to  Ignatyuk's 
prescription \cite{Igna}, 
\be 
a_{ground} = a (1 + f(E^*) \frac{\Delta E_{Shell}}{E^*}),
\ee
with
\be
f(E^*) = 1-e^{- \frac{E^*}{E_{damp}}}.
\ee

In the previous equation $E^*$ is the excitation energy.
The parameter $E_{damp}$ has been adjusted with measured level 
densities and represent how 
much the shell correction is damped by the excitation energy. Its 
value in the  code is taken 
as $18.5$ MeV. At the saddle point we assume no shell correction.

\subsection{Transmission coefficient}{\label{sstc}}

Assuming as usual that, ${\cal T}^J_{l_b S_b}(b,\epsilon_b) \approx  
{\cal T}_{l_b}(b,\epsilon_b) \equiv  {\cal T}_{l} $, we have to solve 
the Schr\"odinger 
equation,
\be
\left( \frac{d^2}{dr^2} - \frac{l(l+1)}{r^2} - \frac{2 \mu}{\hbar^2} 
\left[ V_{Coul}(r) + V_{Nuc}(r) \right]  \right) \psi(r) = 0
\ee
from $r=0$ to infinity in order to exploit the 
asymptotical properties of the wave function. At very large values of 
$r$ we can calculate the $S_l$ matrix and then obtain 
the transmission coefficient ${\cal T}_l = 1- |S_l|^2 $. 
In this equation $V_{coul}$ and  $V_{nuc}$ represent 
coulomb and  nuclear potentials,  respectively.  

At infinity for a given angular 
momentum $l$ the general solution of the Schr\"odinger equation reads
\be
\psi(r) \propto {\cal H}^{(-)}_l(\eta , kr) - S  {\cal 
H}^{(+)}_l(\eta ,kr),
\ee
with 
\beqn
 {\cal H}^{(+)}_l(\eta,kr) &=&  {\cal G}_l(\eta,kr) + i  {\cal 
F}_l(\eta,kr),\\
 {\cal H}^{(-)}_l(\eta,kr) &=&  {\cal G}_l(\eta,kr) - i  {\cal 
F}_l(\eta,kr).
\eeqn
In those equations ${\cal F}_l(\eta,kr) $ and ${\cal G}_l(\eta,kr) $ 
represent the regular and irregular Coulomb wave 
functions,  respectively. 
By using the values of the function $\psi (r)$ at two closes values 
$r_1$ and $r_0$ with 
$r_1 > r_0 \gg 0$ we can obtain the value of the $S_l$ matrix by the 
following formula,
\be
S_l = \frac{ {\cal H}^{(-)}_l(\eta,kr_0)\psi(r_1) - {\cal 
H}^{(-)}_l(\eta,kr_1)\psi(r_0) }{{\cal H}^{(+)}_l(\eta,kr_0)\psi(r_1) 
- {\cal H}^{(+)}_l(\eta,kr_1)\psi(r_0)   }.
\ee
Then we can easily obtain the value of the transmission 
coefficient ${\cal T}_l $.

\begin{table}[H]
\begin{center}
\begin{math}
\begin{array}{|l|c|c|}	
\hline 
 & Neutrons & Charged ~ particles \\ \hline
U_R & 47-0.27E_{lab} & 54-0.32E_{lab}+0.4 
\frac{Z}{A^{\frac{1}{3}}} + 24 \frac{A-2Z}{A}   \\ \hline
R_R & (1.332-7.6 10^{-4} A + 4 10^{-6} A^2 - 8 10^{-9} 
A^3)A^{\frac{1}{3}}  & 1.17 A^{\frac{1}{3}} \\ \hline
a_R & 0.66 & 0.75 \\ \hline
U_{SO} & 7 & 6.2 \\ \hline
R_{SO} & (1.266-3.7 10^{-4} A + 2 10^{-6} A^2 - 4 10^{-9} 
A^3)A^{\frac{1}{3}}    &  1.01A^{\frac{1}{3}}   \\ \hline
a_{SO} & 0.48 & 0.75 \\ \hline
W_I & 0.22E_{lab} - 2.7  & 0.22E_{lab} - 2.7     \\ \hline
R_I & (1.266-3.7 10^{-4} A + 2 10^{-6} A^2 - 4 10^{-9} 
A^3)A^{\frac{1}{3}} & 1.32 A^{\frac{1}{3}}  \\ \hline
a_I  & 0.48 & 0.51 + 0.7\frac{A-2Z}{A}   \\ \hline
W_{SF} & 9.52-0.053E_{lab} & 11.8-0.25E_{lab}+12\frac{A-2Z}{A} \\ \hline
R_{SF} & (1.266-3.7 10^{-4} A + 2 10^{-6} A^2 - 4 10^{-9} 
A^3)A^{\frac{1}{3}} &  1.32 A^{\frac{1}{3}}   \\ \hline
a_{SF} & 0.48 & 0.51 + 0.7\frac{A-2Z}{A}   \\ \hline 
\end{array} 
\end{math}

    \caption{Values of the parameters entering the optical nuclear 
    potential}
    \label{optipot}	
\end{center}
\end{table}

To solve this problem in the KEWPIE code we make all the 
calculation numerically, using the most efficient numerical 
method for evaluation of the wave function, namely the modified 
method of Noumerov \cite{Nou}.

The nuclear optical potential depends on the type of the evaporated 
particle. 
Its real and imaginary parts are given as follows, respectively,
\beqn
U_{Nuc}(r) &=& U_R f(r,R_R,a_R) + \frac{U_{SO} J 
\lambdabar_{\pi}^2}{r} 
\frac{d f(r,R_{SO},a_{SO})}{dr},\\
W_{Nuc}(r) &=& - W_I f(r,R_I,a_I) + W_{SF}4 a_I \frac{d 
f(r,R_{SF},a_{SF})}{dr},
\eeqn
with 
\be
f(r,R,a) = \frac{1}{1+e^{\frac{r-R}{A}}}.
\ee
In these equations $J$ represents the angular momentum and 
$\lambdabar_{\pi}$ the pion Compton wavelength. The other 
parameters are listed in table \ref{optipot}.  We use the values 
given in
Ref. \cite{Wiho} and Ref. \cite{Begr} for neutrons and charged 
particles, 
respectively.

Those parameters have been obtained by fitting the angular 
distributions of proton and neutron scattering. There are no common 
parametrisation for both neutron and protons. We assume that all 
charged particles have the same behaviour as protons, 
with respect to optical potential.

\section{Validation and applications}
\subsection{Statistical observables}\label{stao}

In order to test the code, its results are compared to data
  coming from various experiments related
 to the production of heavy and super-heavy elements. Those 
comparisons are done in two steps.  First we will look at the results 
of the 
evaporation process from a fusion cross section calculated with a 
full 
proximity potential, where the fusion is supposed not to be
 affected by fusion hindrance. 
Next, we will look at the evaporation residue cross sections
 of some superheavy  elements, where the fusion is calculated 
by the two step model  which takes into account the fusion 
hindrance in a realistic way.

\begin{figure}
\epsfxsize=10cm
$$
\epsfbox{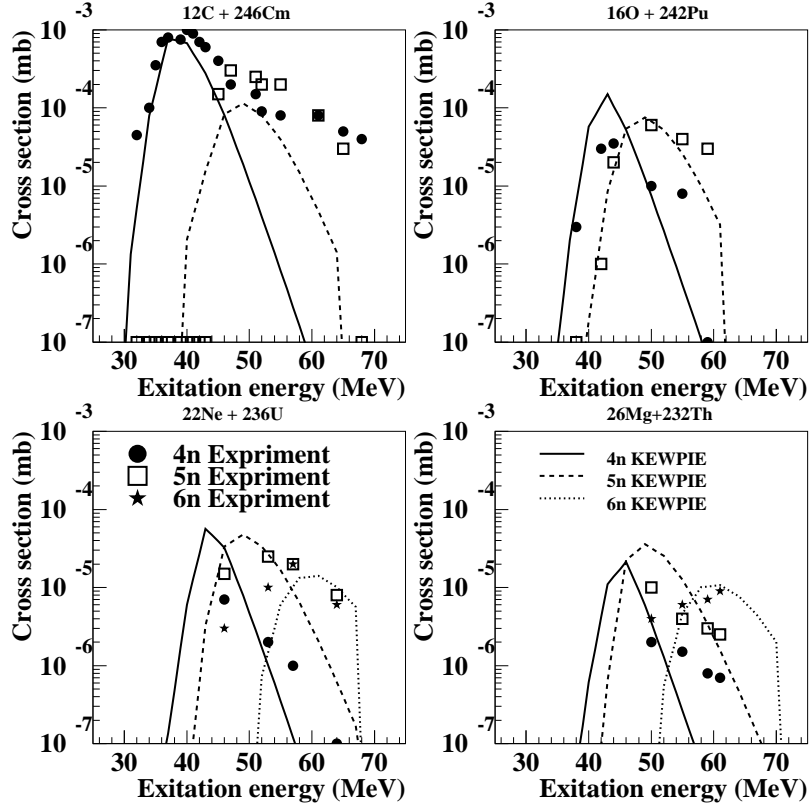}
$$
\caption{\it Cross sections of the residues coming from the 4n,5n and 
6n disintegration processes for $^{12}C + ^{246}Cm $, $^{16}O + 
^{242}Pu $, $^{22}Ne + ^{236}U $ and $^{26}Mg + ^{232}Th $  
reactions. The symbols are representing the experimental data and  
the curves are the results of the calculations with the KEWPIE code.
\label{fig1}}
\end{figure}

We look at the $xn$ residue cross section (residue after evaporation 
of $x$ neutrons)  for various combinations of projectiles and targets 
chosen so as to form the same compound nucleus  $^{258}_{156} No^{102}$.
In fig. \ref{fig1} are represented the cross sections of evaporation 
residues of the $xn$ reactions as a function of the excitation energy 
for the $^{12}C + ^{246}Cm $, $^{16}O + ^{242}Pu $, $^{22}Ne + 
^{236}U $ and $^{26}Mg + ^{232}Th $ reactions.  The experimental data 
are taken from the reference \cite{Sikk,Mik,And} for the  $^{12}C + 
^{246}Cm $ ,  
  $^{16}O + ^{242}Pu $ and  $^{22}Ne + ^{236}U $ and $^{26}Mg + 
^{232}Th $ 
reactions, respectively. In this figure the symbols represent the 
experimental data and the lines the results obtained  by the code 
KEWPIE. 

 In the present case, as stated above, the fusion probability and 
thus fusion 
cross sections, are  calculated 
by using a full proximity potential \cite{Blo} as nuclear attractive 
potential. The fusion hindrance is not taken into account, because  
the products of the charges of the projectiles and targets are lower 
than $1500-2000$, which is the empirical value above  which the 
fusion hindrance appears \cite{Bjsw,Rei2}. 
The calculations of the fusion probability are the same as the 
corresponding part of the HIVAP code \cite{Rei1}. 

\begin{figure}[htbp]
\epsfxsize=10cm
$$
\epsfbox{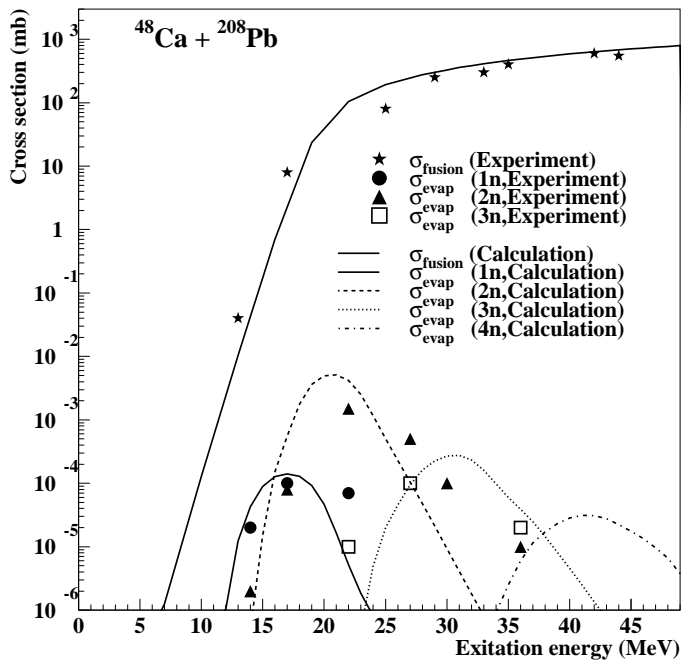}
$$
\caption{\it Cross sections for the fusion and 1n, 2n, 3n and 4n 
evaporation processes from the reaction $^{48}Ca + ^{208}Pb$. The 
symbols represent the experimental data and  the curves the results 
of the KEWPIE code.
\label{fig2}}
\end{figure}

KEWPIE code is designed in such a way that there are as few free 
parameters as possible. The only unconstrained value is the friction 
coefficient $\beta$ for the Kramers factor (see equation 
(\ref{krame})). The value of $\beta$ is taken as  $5~10^{20} s^{-1}$. 
Setting the $\beta$ to $5~10^{21}$ (consistent with the one body wall 
and window formula) reduces too much the fission and this lead to an 
increase of the residue cross section by roughly one order of 
magnitude which obviously overestimates the data as shown in figure 
\ref{fig2}. No other free parameters are used.  It might suggest 
that the fusion hindrance exists somehow in those systems.

The code takes into account the competition between the evaporation 
of neutrons, protons and alpha particles as well as gamma emission, 
and thus a large number of paths in the disintegration process. 
However, the charged particle evaporations can be neglected without 
changing drastically the results. The physical justification  comes 
from  the huge Coulomb barrier that makes the probability of  charged 
particles emissions very low. That approximation could be used to 
save calculation time. 

The figure \ref{fig1} shows that an overall global agreement is 
obtained even with the strict conditions we have put and without 
adjusting free parameters that are usually employed in statistical 
analyses of data.

To complete the evaluation of the code we looked at a reaction with a 
more symmetric entrance channel to produce nobelium isotopes: 
$^{48}Ca + ^{208}Pb \rightarrow ^{256}_{154} No^{102}$. The product 
of the charges of the target and the projectile ($Z_p \times Z_t = 
1640$)  indicates that the fusion process described by the transmission 
of the barrier with the full proximity potential is still enough to 
give a good evaluation of the fusion cross-section. The residue cross 
section of the $xn$  evaporation processes  are shown on figure 
\ref{fig2}.

As in the previous calculation shown in  figure \ref{fig1}, $\beta$ 
is set to $5.10^{20} s^{-1}$. As we can see in the figure \ref{fig2} 
the calculation well reproduces the experimental data as a whole. If 
we look more in details, we can see that the peaks of the 
distribution are always located at the suitable positions. The main 
discrepancies come from the widths of the peaks. The calculated 
distributions are a little narrower than those of experiment. This 
small disagreements would come both from the experimental and the 
theoretical accuracies.  In experiment there is  a kind of averaging 
over a range of excitation energy of the residue cross sections due 
to the resolution of the detection set-up and the lose of energy of 
the beam in the target, those effects making the distributions 
broader. On the theoretical side, the problem of the width of the 
evaporation can come from the assumption made on the duration of one 
decay step. In the usual calculation code the time between two steps is 
suppose to be infinite. But if we caclulate each step with a finite time,
 the distributions become wider due to the progressive 
feeding of some nuclei. Moreover the inclusion of a transient time 
should especially affect the high energy part of the residue spectrum 
and make the whole distribution wider. 

We can conclude  that in the reaction of production of the nobelium 
isotopes the fusion cross sections calculated with the full proximity 
potential give rise to an  overall good agreement with the data, 
combined with the present statistical decay calculation.

\subsection{Dynamical observable}

In order to test the dynamical description of the code, a suitable 
observable would be the fission time which could be easily calculated 
by using equation (\ref{fisti}). The case of 
$^{238}U$ is particularly interesting because its fission barrier has 
roughly the same value as the neutron binding energy. Therefore, 
competition between the two channels occurs all along the 
disintegration chain and gives rise to very long fission times. The 
fission times were measured directly by crystal blocking techniques 
which are nuclear model independent \cite{Mor}.

\begin{figure}[htbp]
\epsfxsize=10cm
$$
\epsfbox{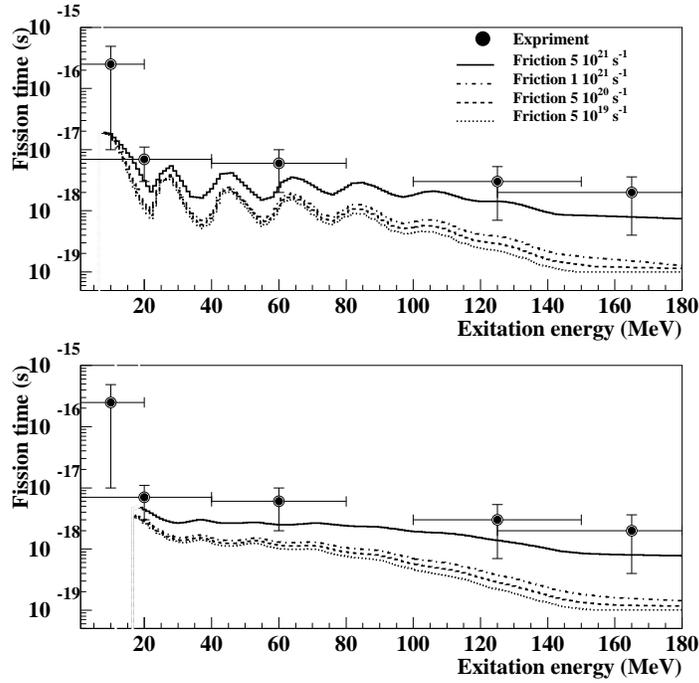}
$$
\caption{\it Fission time of $^{238}U$  as a function of the 
excitation energy. The upper panel represents the  calculation done 
with a step of excitation energy of 3 MeV with  angular momentum of 
the compound nucleus $J = 1,5 $ and $9 \hbar$. The lower panel shows 
an averaging of the calculation on a range of 30 MeV. The curves 
represent the calculation done with various values of the friction 
factor $\beta$. The symbol represent the experimental data.  
 \label{fig4} }
\end{figure}

Figure \ref{fig4} shows a comparison between theoretical calculation 
done with KEWPIE code and data from reference \cite{Mor}. We take 
only into account the competition between fission and neuton 
evaporation. Moreover, the emission of gamma ray are taken 
into account in this process since its cooling effect at the end of 
the decay, when low exitation energies are reached, affects the 
fission time. Those approximations are done in order to save 
calculation time but are resonable with respect to the evaluation of 
the fission time. In this experiment the angular momentum of the 
fissioning $^{238}U$ is suppose to be lower than $10 \hbar$. In the 
calculation three values of the angular momentum $1 \hbar  ,5\hbar$ 
and $9\hbar$ are  taken.

The upper part of the figure \ref{fig4} shows the fission time 
(define by equation \ref{fisti}) as a 
function of the excitation energy. By a careful study of the output 
of the calculation, we noticed that the average value of the fission 
time is affected by very slow fission events at the  end of the decay 
chain.
The calculated results show some oscillations that are damped at high 
energy.  The origin of these oscillations is mainly a threshold 
effect : to emit $x$ neutrons a certain value of the energy must be 
reached. When the energy reachs the suitable value, the fission time 
increases due to the stronger competition between fission and 
evaporation. Since each neutron emitted carries some kinetic energy, 
proportionally to the temperature of the nucleus under consideration 
(square root of the energy), the period of the oscillations increases 
with excitation energy. However peaks are separated by roughly 20 
MeV, which is a too large value compared with the binding energy of a 
neutron.  If we remove  the pairing energy from the mass of the nuclei, 
new peaks appear between the peaks shown on figure 
\ref{fig4}.  This succession of peaks with a periode of 
approximatively 10 MeV can now be easily understood as a threshold 
effect. Thus the oscillations and their behavior come from the 
combined effects of energy threshold and pairing energy. 
We notice also that as we go to higher energy, the amplitude of the 
oscillation decreases. The damping of those oscillations can be 
explained by the increase of the number of steps of evaporation at 
higher energy. This conduces to a boarding of the energy 
distribution. Thus the threshold effect becomes less sensitive and 
the amplitude of the oscillation is reduced. At low excitation energy 
the disagreement between the calculations and the experimental data 
looks more important. This can be attributed to the quantum effect 
that persists at such low temperature of the compound nucleus. This 
hypothesis is confirmed by the experimental observation of asymmetric 
fission below 50 MeV in this experiment.

We see that the experimental data present no oscillation, but have 
very large error bars. The huge size of the error bars would be  the 
main argument to explain the absence of oscillations. Actually, if we 
plot the same calculation as in the upper panel but averaged on a 
range of 30 MeV (range  comparable to the mean value of the error 
bars) shown in the lower panel of the figure 4, we can see that the 
oscillations roughly disappear.  

The agreement between the experimental data and the calculation are 
reasonably good. The experimental  value and the calculation are all 
in agreement within a range of roughly one order of magnitude.  The 
evolution of the calculated curves as a function of the friction 
parameter $\beta$ is rather strong  when we go from $\beta = 1 ~ 
10^{21} s^{-1}$ to $\beta =5 ~ 10^{21} s^{-1}$ this is due to the 
Kramers factor (see equation (\ref{krame})) as describe in the 
section \ref{fig1}. It seems that the friction value of  $5 ~ 10^{21} 
s^{-1}$ is the best to reproduce experimental data. However, we have 
to be careful on some points to give definitive conclusion: information 
is missing on angular momentum 
distribution.  The reliability of the conclusion could  probably be 
improved by having more information on the distribution of angular 
momentum as a function of the excitation energy.

We should notice that in those calculations the value of the reduce 
friction coefficient $\beta$ that gives the best agreement with the 
data is $5 ~ 10^{21} s^{-1}$ and in the section \ref{stao} 
the value of $\beta$ is set to $5 ~ 10^{20} s^{-1}$. This difference 
in the suitable value of $\beta$  is a rather delicate problem which 
goes beyond the scope of this paper. Detailed analysis will be published elsewhere.

\section{Conclusion}

The KEWPIE code has been designed as a disintegration code dedicated 
specifically to the study of superheavy elements. Its conception is 
based on the idea of minimising the number of free parameters. So the 
only free parameter that is adjusted in the calculation is the 
reduced friction coefficient $\beta$. We have shown that the results 
of the calculation with such a high level of constraint of only one 
adjustable parameter are rather successful. The quality of the 
agreement is as good as other evaporation codes that are including 
more adjustable parameters. The results of the calculation on the 
superheavy elements  show that 
the shell correction energy is a key physical quantity for 
understanding and reproducing the residue cross sections.

The architecture of the program as a full cascade is an innovation 
which permits us to include the dynamics of the reaction. Thus, the 
evolution during the decay can be calculated. This time dependence 
permits to calculate fission time dynamically. The results of the 
calculations appear to be in quite a good agreement with the 
experimental data. 

The KEWPIE program permits to study the disintegration of nucleus and 
to accommodate both a statistical and a dynamical frameworks. 
  Moreover the way of 
programming allows us to implement new physical effects in the code 
without affecting the whole structure.

\begin{acknowledgments}
The authors thank JSPS for support (contracts P-01741, 1340278 and S-02727). 
Two of us thank the Yukawa Institute for its warm hospitality. 

\end{acknowledgments}

\end{document}